%% file: main-arxiv.tex
\documentclass[sigconf]{acmart}
\usepackage{subcaption}

\AtBeginDocument{%
  \providecommand\BibTeX{{%
    \normalfont B\kern-0.5em{\scshape i\kern-0.25em b}\kern-0.8em\TeX}}}

\setcopyright{none}
\copyrightyear{2022}
\acmYear{2022}
\acmDOI{XXXXXXX.XXXXXXX}
\renewcommand\footnotetextcopyrightpermission[1]{}
\usepackage{algorithmic}

\newcommand{\etal}{\textit{et al.}}

\usepackage{xcolor}

\renewcommand{\vec}[1]{\boldsymbol{#1}}
\newcommand{\NN}{\mathbb{N}}

\newcommand{\CommitHistory}{\mathcal{CH}}
\newcommand{\EventSet}{\mathcal{E}}
\newcommand{\bitcoin}{\textit{BC}}
\newcommand{\Bitcoin}{\bitcoin}
\newcommand{\altcoin}{\chi}
\newcommand{\eventmap}{\phi}

\newcommand{\nbcc}{\textrm{nbcc}}
\newcommand{\nbct}{\textrm{nbct}}

\newcommand{\ours}{\texttt{GitWatch}}
\newcommand{\heuristA}{PCF}
\newcommand{\heuristB}{PEF}
\newcommand{\heuristC}{PTF}
\newcommand{\delayA}{\Delta^{\textrm{PCF}}}
\newcommand{\delayB}{\Delta^{\textrm{PEF}}}
\newcommand{\delayC}{\Delta^{\textrm{PTF}}}

\begin{document}

\title{Estimating Patch Propagation Times across (Blockchain) Forks}

\author{Sebastien Andreina}
\affiliation{%
  \institution{NEC Labs Europe}
  \city{Heidelberg}
  \country{Germany}}
\email{sebastien.andreina@neclab.eu}

\author{Lorenzo Alluminio}
\affiliation{
  \institution{NEC Labs Europe}
  \city{Heidelberg}
  \country{Germany}}
\email{lorenzo.alluminio@neclab.eu}

\author{Giorgia Azzurra Marson}
\affiliation{
  \institution{NEC Labs Europe}
  \city{Heidelberg}
  \country{Germany}}
\email{giorgia.marson@neclab.eu}

\author{Ghassan Karame}
\affiliation{
  \institution{Ruhr University Bochum}
  \city{Bochum}
  \country{Germany}}
\email{ghassan.karame@rub.de}

\renewcommand{\shortauthors}{Andreina et al.}

\begin{abstract}
  The wide success of Bitcoin has led to a huge surge of alternative cryptocurrencies (altcoins). 
  Most altcoins essentially fork Bitcoin's code with minor modifications, such as the number of coins to be minted, the block size, and the block generation time. As such, they are often deemed identical to Bitcoin in terms of security, robustness, and maturity.

  In this paper, we show that this common conception is misleading. By mining data retrieved from the GitHub repositories of various altcoin projects, we estimate the time it took to propagate relevant patches from Bitcoin to the altcoins.
  We find that, while the Bitcoin development community is quite active in fixing security flaws of Bitcoin's code base, forked cryptocurrencies are not as rigorous in patching the same vulnerabilities (inherited from Bitcoin).
  In some cases, we observe that even critical vulnerabilities, discovered and fixed within the Bitcoin community, have been addressed by the altcoins tens of months after disclosure. Besides raising awareness of this problem, our work aims to motivate the need for a proper responsible disclosure of vulnerabilities to all forked chains prior to reporting them publicly.
\end{abstract}

\maketitle
\pagestyle{plain}

\input{introduction}

\input{background}

\input{model}

\input{methodology}

\input{analysis}
\input{output}
\input{related-work}

\input{conclusion}

\section*{Acknowledgments}
This work was partially funded by the Deutsche Forschungs-gemeinschaft (DFG, German Research Foundation) under Germany’s Excellence Strategy - EXC 2092 CASA - 390781972 and the European Union (INCODE, Grant Agreement No 101093069). Views and opinions expressed are however those of the author(s) only and do not necessarily reflect those of the European Union. Neither the European Union nor the granting authority can be held responsible for them.

\bibliographystyle{ACM-Reference-Format}
\bibliography{references}

\end{document}

%% file: introduction.tex
\section{Introduction}
\label{sec:introduction}

The wide success of Bitcoin has led to an explosion in the number of so-called ``altcoins'', i.e., cryptocurrencies designed as a fork of the Bitcoin-core code base.
Although altcoins share---to a large extent---the same technical foundations of Bitcoin, they feature minor differences to Bitcoin.
For instance, some altcoins exhibit a different block-generation time (e.g., Dogecoin and Litecoin), use a different hash function (e.g., Litecoin and Namecoin), or impose a different limit on the supply amount (e.g., Dogecoin, Litecoin)~\cite{DBLP:conf/ccs/GervaisKWGRC16}.
Often users and practitioners assume that altcoins offer a similar level of security and stability as Bitcoin-core. Indeed, the predominant choice for users to adopt a given altcoin over Bitcoin seems to hinge on the perceived profitability of mining that altcoin~\cite{cointracker} rather than its actual security and maintenance effort of the developing team.

Bitcoin (and many of its descendants) have been found vulnerable to a wide variety of attacks.
Gervais~\etal~\cite{DBLP:conf/ccs/GervaisKWGRC16} proposed the first quantitative framework to analyze the security and performance of proof-of-work blockchains, hinting that some altcoins might offer weaker security compared to Bitcoin, owing to the various, often ad-hoc parameters that they adopt.
Thanks to its strong developing team, the Bitcoin-core software is routinely monitored and promptly patched.
However, whether altcoins are also as proactive in fixing disclosed vulnerabilities remains unclear.

In this paper, 
we investigate the robustness and stability
of altcoins from the perspective of code maintenance and patching.
More specifically, we identify vulnerabilities reported in Bitcoin, and study how their patches are propagated through various altcoins.
Our approach relies on the inspection of GitHub repositories of popular cryptocurrencies, to identify relevant bugs and corresponding patches in the commit history of GitHub-hosted altcoin projects.
Concretely, we aim to estimate the time it took to propagate patches from Bitcoin to various altcoins, i.e., to determine whether and how quickly altcoins address disclosed security vulnerabilities.

However, retrieving detailed timing information associated to code changes in GitHub, especially in the context of Bitcoin forks, emerges as a challenging task.
The reason is that most patches are taken directly from the Bitcoin-core repository and applied to a Bitcoin fork via a~\texttt{rebase} operation which only exposes the timestamp of the original patch (applied to Bitcoin) and not the actual time when the patch was ported (to the altcoin)~\cite{berger}.
When patches are ported via rebasing,
it is not possible to determine patch propagation times from the information recorded in GitHub.
Indeed, every rebasing replaces all current commits in a fork by new commits that apply the same changes but with an updated and fresh timestamp.
In addition, the original commits are no longer referenced after~\texttt{rebase} occurs.
As Git prunes unreferenced commits periodically, the timestamps associated to a given patch are lost with every subsequent~\texttt{rebase} invocation.

To overcome this challenge we propose {\ours}, a tool to measure patch propagation times in Git-hosted forked projects even in the case of patches ported using the~\texttt{rebase} command.
{\ours} leverages GitHub's event API and GH archive to estimate the time when a given patch is applied to a forked project.
{\ours} design relies on the following observation:
although GitHub follows the same practices as Git, pruning unreferenced commits, internally it keeps a log of metadata information for all commits that ever existed.
Importantly, this information can be retrieved through GitHub's API, as long as one can reference the relevant commits.
Technically, {\ours} makes use of the GH archive service to generate a list of all commits pertaining to a given GitHub project, before querying GitHub to retrieve the commits' metadata.
For each forked project under scrutiny, we harvest all events from GitHub using GH archive, and we reconstruct the full graph of commit operations---including unreferenced commits related to the sought patches.
By monitoring all past GitHub commits instead of inspecting the sole commit history of the fork, we are able to estimate realistic patching times despite rebasing.

We validate~{\ours} in the context of analyzing the security of altcoin projects, and we study the time to propagate patches from Bitcoin to various altcoin projects.
We consider~47 patches comprising 11~vulnerabilities reported in academic papers, 23~Bitcoin's Common Vulnerabilities and Exposures (CVEs), 3 major CVEs in libraries used by Bitcoin, 3 Bitcoin improvement proposals and 7 major bugs found on the GitHub repository with tags related to the peer-to-peer network, covering crucial vulnerabilities reported in the last decade (see Table~\ref{tab:vulnerability:list}).
Our study investigates the security of five altcoins, which we selected among existing GitHub-based open-source forks of Bitcoin to ensure diversity in terms of market cap, popularity, and vision.

Our results (cf.~Section~\ref{sec:methodology}) indicate that, for all of the selected altcoins, most patches have been applied with considerable delay compared to the release time of the Bitcoin fix, thereby leaving users running vulnerable software, that could be exploited by attackers, for several months or even years.
In contrast to common belief, these findings suggest that Bitcoin forks---due to their less reactive software maintenance---may contain published vulnerabilities in their code base for several months, up to a few years for the worst case, showing therefore weaker robustness and security compared to Bitcoin Core.

We hope that our results motivate altcoin developers to promptly react to disclosed patches; another important purpose of our work is to motivate the need for a proper responsible disclosure of vulnerabilities to all altcoins
prior to any publication of the vulnerability.

Notice that {\ours} is applicable to other forks beyond cryptocurrency projects, as it provides  
a workable means to extract reliable timing information of selective patches when these patches are ported via rebasing public repositories.
For instance, {\ours} can be applied for the
analysis of patch-propagation times in highly forked open source projects such as the Linux kernel~\cite{linuxkernel}, with more than 42000 forks, or the Bootstrap library~\cite{bootstrap}, with more than 76000 forks.

%% file: background.tex
\section{Background \& Problem}
\label{sec:background}

In this section, we overview preliminary concepts about Bitcoin, altcoins, and Git.

\subsection{Bitcoin and Altcoins}
Bitcoin is the first peer-to-peer system that implements a fully-decentralized cryptocurrency~\cite{btc_paper}.
The Bitcoin client software, called ``Bitcoin Core'', is maintained by a large open source developer community and is regularly updated, reviewed, patched and tested~\cite{bitcoincore}.

A few years after the inception of Bitcoin in 2009, new cryptocurrencies were created and have since then mushroomed---at the time of writing there are more than~11'000 cryptocurrencies~\cite{coinmarketcap}.
The reasons for introducing alternative cryptocurrencies are diverse, e.g., to improve Bitcoin's design, to support additional features, or to customize the protocol to specific applications.
To this end, many projects are directly derived from Bitcoin by forking the official code base hosted on GitHub.

In this paper, we refer to alternative cryptocurrencies that are based on Bitcoin Core's software as \emph{altcoins}. Prominent altcoins are Dash~\cite{dash}, initially renamed as ``Darkcoin'' for its intense adoption in dark-net markets;
Digibyte~\cite{digibyte}, a cryptocurrency that was advertised by its creators for improving functionality via real-time difficulty adjustment, nearly instantaneous transactions with~15 seconds confirmation, and enhanced security compared to Bitcoin;
Monacoin~\cite{monacoin}, a cryptocurrency created by an anonymous community in Japan that aimed at developing a national cryptocurrency payment system, as well as
Litecoin~\cite{litecoin} and Dogecoin~\cite{dogecoin}, two famous cryptocurrencies that have emerged among the most popular first-generation derivatives of Bitcoin.

\subsection{Lack of Reliable Patching Timestamps in Git}
\label{sec:git}

Git is a distributed version control tool~\cite{git}. Git resources are widely adopted for the development of collaborative, open-source projects (including cryptocurrencies), as they enable the various collaborators to easily and concurrently handle different versions of the source code.
A Git-hosted project can be managed through various operations, providing users with read and write access to the files stored in the project repository.
More concretely, a Git repository records the history of changes made by users in the form of a sequence of \emph{snapshots}, so that one can inspect the repository's content at any point in time by retrieving the corresponding snapshot. 
To join a collaborative project, a user starts by cloning the content of the project's repository and synchronizing its local copy with the remote version (by invoking~\texttt{pull}). This operation allows incorporating changes made by other users and obtaining the sequence of existing snapshot---akin to a~``read'' operation.
In order to~``write'', a user can~\texttt{commit} changes to its local directory and then~\texttt{push} the commits to the remote repository.

Cloned projects typically port software patches (that have been applied to the parent project) via~\texttt{rebase} operations. Unfortunately, every~\texttt{rebase} invocation  modifies the history of the fork's repository, in particular altering the timestamps of all commits re-applied to the fork, with the effect of erasing the timestamps of all previously ported patches.
Therefore, retrieving accurate patch propagation times directly from the commit history of a fork is not possible in Git. This clearly prevents any reliable study of patch-propagation timing for cloned code.

%% file: model.tex
\section{Measuring Patch Propagation Times in Git}
\label{sec:model}

In this section, we study the problem of analyzing the propagation times of patches in Git across forked projects (i.e., the time it takes a patch to be ported from the main project to the forked project). We begin with formalizing the various Git operations that are relevant to software patches, then we discuss in detail the effect of~\texttt{rebase} operations on commit timestamps.
Finally, we propose three heuristics that can be used to estimate the patch-propagation time for GitHub-hosted forked projects.

\subsection{Git Operations}
\label{sec:git:ops}

\paragraph{Commit}
This command allows tracking changes made to one or more files in the repository.
We define a commit as a pair~$C = (M,D)$ of \emph{metadata} and~\emph{data}, containing context information about the commit and the actual changes made by that commit, respectively.
Below, we highlight the metadata fields that are most relevant to describe our methodology in the next section:
\begin{align}
    h &: \text{commit hash (or commit ID)},\\
    p &: \text{parent commit},\\
    a &: \text{author},\\
    c &: \text{committer},\\
    t_a &: \text{author timestamp},\\
    t_c &: \text{committer timestamp},
\end{align}
where~$h$ is a cryptographic hash over the changes~$D$ along with the remaining metadata, i.e.:
\begin{equation}
    h = H(p,a,c,t_a,t_c,D),
\end{equation}
and it uniquely identifies a given commit;
the parent~$p$ is a reference to the previous commit;
the author~$a$ denotes the author of the changes introduced in the commit,
while the committer~$c$ denotes the user who made the latest changes to that commit;
the author timestamp~$t_a$ indicates when (date and time) the original commit operation was performed by the author,
and the committer timestamp~$t_c$ records when the latest change was made (to that commit) by the committer.
Whenever the commit is amended, the committer is replaced with the user who modified the commit and similarly the commit timestamp is updated to the current time.
Metadata information is essential for examining the \emph{history} of a repository.

Git allows associating \emph{tags} to commit operations (e.g., to mark released versions of software).
Given a commit~$C$ with commit hash~$h$, a tag referencing~$C$ is defined by a tuple~$\tau = (\lambda,t,h)$ where~$\lambda$ is a human-readable label (e.g., the version number of the release) and~$t$ is the timestamp of the tag.

\paragraph{Push}
A user can apply its local changes to a remote repository by invoking~\texttt{push}. This operation applies all (new) local commits to the remote repository.
Each such commit corresponds to an updated snapshot of the repository's content.
A ``batch'' of commits pushed by a given user forms a sequence~$(C_1,\dots,C_m)$ defined implicitly by the references to parent commits. Formally, if~$u$ denotes the user who created and pushed the commits, for~$i = 1,\dots,m$, we have:
\begin{equation}
    C_i.a = C_i.c = u
    \quad\land\quad
    C_i.t_a = C_i.t_c,
\end{equation}
i.e., author and committer coincide with the user who pushes the commit,
and so do the author timestamp and committer timestamps.
In addition, for~$i = 2,\dots, m$, it holds:
\begin{equation}
    C_i.p = C_{i-1}.h
    \quad\land\quad
    C_i.t_c > C_{i-1}.t_c,
\end{equation}
i.e., every commit, except for the first one in the sequence, has the previous local commit as parent, and the commit timestamps in the sequence progressively increase (i.e., the commit order reflects the chronological order of execution).

Pushing a sequence of commits triggers the addition of new snapshots, typically one per commit~$C_i$, to the history of the repository.
Let~$\CommitHistory$ denote the commit history of the repository, i.e., the collection of commits that have been pushed so far.
Then, pushing~$(C_1,\dots,C_m)$ has the effect of appending the commit sequence to the commit history:
\begin{equation}
    \CommitHistory \stackrel{\texttt{push}}{\longleftarrow} \CommitHistory \parallel \{ C_1,\dots,C_m\} .
\end{equation}
Notice that the commit history is not always a sequence, due to commits pushed concurrently by different users.
For instance, merged commits have the same parent commit, thus forming a directed acyclic graph (rather than a sequence).
In the rest of the paper, to simplify the notation we write the commit history as a sequence.

\paragraph{Fork}
A fork (a.k.a.~\textit{branch})
of an existing repository~$R$ is a repository~$R^\altcoin$ that shares a common history with~$R$---the latter is called the ~\textit{main branch}.
The latest commit that~$R$ and~$R^\altcoin$ have in common is called \emph{base commit}.
Let~$\CommitHistory$ and~$\CommitHistory^\altcoin$ denote the commit histories of~$R$ and~$R^\altcoin$ respectively.
Then there exist~$m, s, r\in \NN$, $r > 0$, such that:
\begin{align}
    \CommitHistory      &= (C_0,\dots,C_m,\dots,C_{m+s})\\
    \CommitHistory^\altcoin &= (C_0,\dots,C_m,C^\altcoin_1,\dots,C^\altcoin_r)
\end{align}
where~$C_m$ denotes the base commit and~$C^\altcoin_1,\dots,C^\altcoin_r$ are the commits diverging from the main branch.
While forks typically proceed independently of their main branch, the developers of a forked project might still need to monitor the evolution of the main branch, e.g., to discover bugs and port patches.

\paragraph{Rebase}
This operation allows integrating changes from the main branch~$R$ (e.g., Bitcoin) to a fork~$R^\altcoin$ (e.g., an altcoin) by re-applying all commits pushed to~$R^\altcoin$
starting from a new base commit in~$R$---hence the term \emph{rebase}. This operation is most often adopted to fetch the latest version of the original repository.
Invoking~\texttt{rebase} effectively ``re-builds'' the changes made in the fork on top of the new base commit, thereby modifying the commit history of the fork.
Namely, suppose the commit histories of the two repositories are as follows:
\begin{align}
    \CommitHistory &= (C_0,\dots,C_m,\dots, C_{m+s}),\\
    \CommitHistory^\altcoin &= (C_0,\dots,C_m,C^\altcoin_1,\dots,C^\altcoin_r).
\end{align}
Then, invoking~$\texttt{rebase}$ on~$R^\altcoin$ with new base commit $C_{m+k}$, for $0 < k\leq s$, has the following effect on the commit history of~$R^\altcoin$:
\begin{align}
    \CommitHistory^\altcoin \stackrel{\texttt{rebase}}{\longleftarrow} (C_0,\dots,C_{m+k},C'^\altcoin_1\dots,C'^\altcoin_r),
\end{align}
where each commit~$C'^\altcoin_{i}$, for $i=1,\dots,r$, is an updated version of the original commit~$C^\altcoin_i$ adapting  the metadata to the new base commit. Concretely, the committed changes (i.e., the data~$D$) remain the same, i.e., for all~$i=1,\dots,r$, it holds:
\begin{equation}
    C'^\altcoin_i.D = C^\altcoin_i.D,
\end{equation}
and the metadata is updated to reflect the replacement of the base commit~$C_m$ with~$C_{m+k}$. This update modifies the first parent commit as follows:
\begin{equation}
    C'^\altcoin_1.p \gets C_{m+k}.h
\end{equation}
which, in turn, triggers a chain reaction and modifies all subsequent parent commits,
i.e., for $i = 2,\dots,r$, we have:
\begin{equation}
    C'^\altcoin_i.p \gets C'^\altcoin_{i-1}.h
\end{equation}
At a lower level, rebasing does not technically replace every commit~$C^\altcoin_i$ with~$C'^\altcoin_i$, it creates a new sequence~$C'^\altcoin_1,\dots,C'^\altcoin_r$ of commits, each with its own fresh commit ID. However, after rebasing the original commits are no longer referenced, hence they become ``dangling'' commits.

Finally, rebasing also has the crucial effect of updating the committer timestamp with the current time, while the author timestamp is preserved:
\begin{equation}
    C'^\altcoin_i.t_c \gets \text{`current time'} \quad\land\quad C'^\altcoin_i.t_a = C^\altcoin_i.t_a.
\end{equation}

\begin{figure}
    \includegraphics[width=\linewidth]{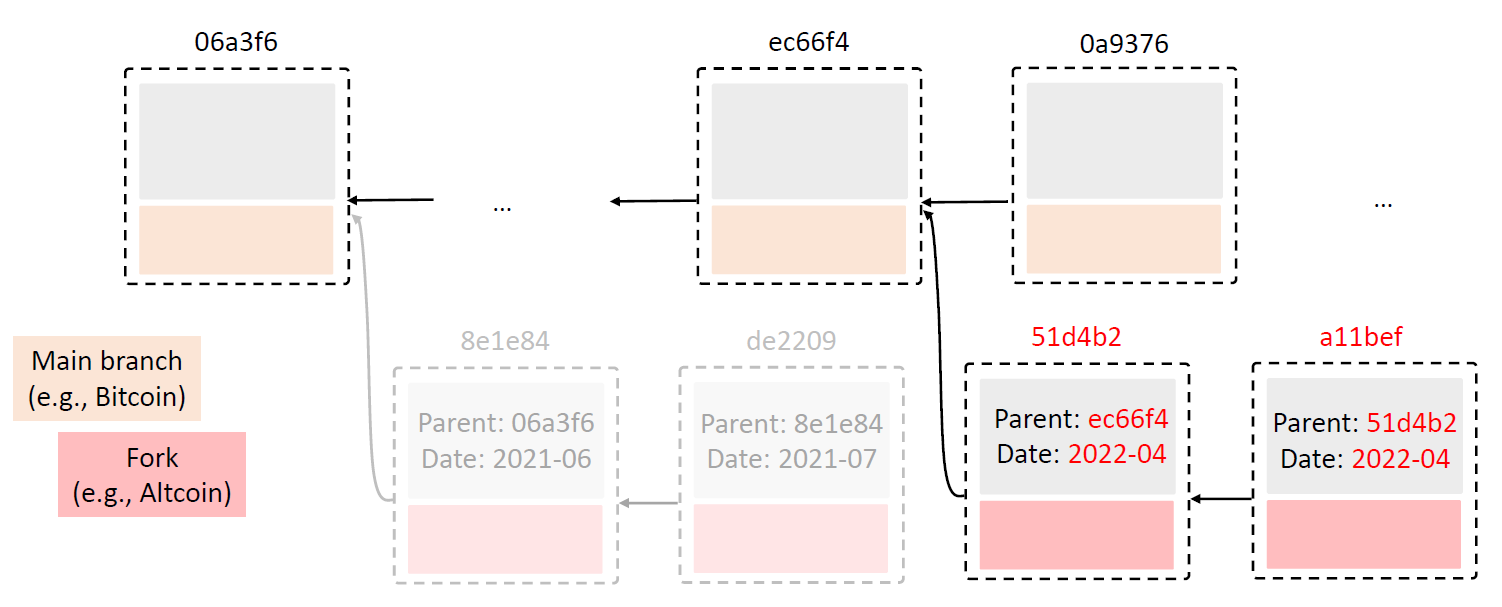}
    \caption{Effect of~\texttt{rebase} on commit metadata: parent commit, committer date, and commit id are modified. Each dotted box represent a commit/snapshot, with metadata (top, gray) and data (bottom, colored), while arrows represent pointers to parent commits.}
    \label{fig:rebase:closeup}
\end{figure}

\paragraph{Rebasing makes timestamps unreliable}
Although recording the rebasing time in the committer timestamp appears as a natural approach to maintain information about relevant events in $\CommitHistory^\altcoin$, it can cause the loss of relevant timing information in the case of multiple~\texttt{rebase} operations being performed on the same repository.
Indeed, every new~\texttt{rebase} invocation preserves the author timestamp~$t_a$ of the original commits,
however, it re-sets all committer timestamps~$t_c$ in the commit history to the current time---thereby overwriting all timestamps of previous~\texttt{rebase} operations.
This behavior is illustrated in Figure~\ref{fig:rebase:closeup}.

After a rebase, the old commits $C^\altcoin_1,\dots,C^\altcoin_r$ become unreferenced and are called ``dangling'' commits. In order to save up space, dangling commits are automatically pruned by Git. However, when a rebase transforms commit~$C_i$ into~$C'_i$, the two commits are factually different due to their differences in the metadata and are therefore initially both accessible via their respective commit ID. Assuming no pruning, this observation provides us with a strategy to retrieve the timestamp of rebases: by listing all the different versions of a commit~$C_i$ and their respective committer timestamp. Our methodology described in Section~\ref{sec:methodology} is based on this intuition to estimate the timing of rebases, yet it is compatible with the pruning of dangling commits.

\subsection{Extracting timing information from GitHub}

GitHub generates events for all operations on public repositories that can be subscribed to.
To extract meaningful information about patch propagation time---even when the patch is applied via rebasing---we rely on two main resources: GitHub's event API and GH archive. GH archive~\cite{GHarchive} is an openly accessible service that provides the history of all events of GitHub since 2011.
We observe that while rebases create dangling commits that are not retrieved when cloning, these commits can still be queried through GitHub's event API by requesting the corresponding hash.

\paragraph{GitHub events.}
Events are associated to one or more commits---e.g., a~\texttt{push} operation triggers a corresponding event associated to the list of commits~$(C_1,\dots,C_m)$ pushed by the author.
We define an event~$e$ as a tuple~$e = (\vec{C_e}, t_e)$, where $\vec{C_e} = (C_1,\dots,C_m)$ is the list of commits associated to the event, and~$t_e$ is the event timestamp recorded in GitHub.
We denote by~$\eventmap$ the mapping from commits to events, i.e., for an event $e = (\vec{C_e},t_e)$, we define~$\eventmap(C_i) := e$ for all~$i=1,\dots,m$.
With a slight abuse of notation, we write~$C \in e$ to indicate that commit~$C$ belongs to the sequence of commits~$\vec{C_e}$ associated to event~$e$.
Using this notation, we have $C \in e \Longleftrightarrow \eventmap(C) = e$.

Using GH archive to retrieve all the hashes pertaining to commits of a given GitHub project, we are able to reconstruct the full graph of commits by fetching the metadata of those commits through GitHub's API---
including dangling commits that are pruned following rebase operations. We effectively use GH archive to find the commit hashes and then query the metadata of those hashes through GitHub's API.
By inspecting the full tree of Git commits along with their timestamps provided by the GitHub event~API, we are able to estimate the time when a patch was introduced in each of the analyzed altcoins using different heuristics.
We then compare these estimates to the original timestamp of the Bitcoin patch, so that we can derive (an estimate for) the patch propagation time from Bitcoin to the altcoin.

\paragraph{Graph Generation.}
Our methodology to obtain the timestamp of a patch is based on the investigation of dangling commits. To this end, for each altcoin~$\altcoin$ we build a graph of all commits, including dangling commits.
Formally, graph~$\mathcal{G}_\altcoin = (V,E)$ contains all commits~$C$ of altcoin~$\altcoin$ as vertices, while the edges represent the parent to child relationship:
\begin{equation}
	C \in \CommitHistory^\altcoin \Longrightarrow
	C\in V \land (C.p,C) \in E.
\end{equation}
Given~$\mathcal{G}_\altcoin$, we consider the following three heuristics to estimate the patch propagation times.

\paragraph{Patch-commit finder (\heuristA)}
The first heuristic relies on the concept of ``non-Bitcoin child commits'' for a given fork~$\altcoin$.
Intuitively, given~$\mathcal{G}$, we locate a non-Bitcoin commit~$C^*_j$ that contains the patch in its history, and use the committer date of~$C^*_j$ as the patching date. The patch propagation time is then estimated to be the time between the committer timestamp of the located commit and the timestamp of the original commit.

Formally, given a Bitcoin commit~$C_i$, i.e., $C_i \in \CommitHistory^\Bitcoin$, we define the set of non-Bitcoin child commits of~$C_i$ as follows:
\begin{equation}
	\begin{gathered}
		C_j\in \nbcc_{\altcoin}(C_i)
		\Longleftrightarrow C_j \not\in \CommitHistory^\Bitcoin\ \land\ C_j\in \CommitHistory^\altcoin\ \land\\
		\exists(E_1,..,E_n) \in \mathcal{G}_\altcoin\ :\ E_1.\mathtt{from} = C_i \land E_n.\texttt{to} = C_j\ \land\\
		\forall i \in [1,n-1],\ E_i.\texttt{to} = E_{i+1}.\texttt{from}.
	\end{gathered}
\end{equation}

In order to estimate the propagation delay of a given patch commit~$C_i\in \CommitHistory^\Bitcoin$, from Bitcoin to an altcoin~$\altcoin$, {\heuristA} locates a%
\footnote{A commit~$C^*_j$ with this property is not necessarily unique, as multiple commits can have the same committer timestamp.}
commit~$C^*_j \in \CommitHistory^\altcoin$ such that:
\begin{equation}
	\begin{gathered}
		C^*_j \in \nbcc_{\altcoin}(C_i)\ \land\ C^*_j.{t_c} \leq C_j.{t_c}\ \forall C_j\in \nbcc_{\altcoin}.
	\end{gathered}
\end{equation}
Finally, we define the estimated patch propagation delay {$\delayA$} as:
\begin{equation}
	{\delayA} \gets C^*_j.{t_c} - C_i.{t_a}.
\end{equation}

\paragraph{Patch-event finder ({\heuristB})}
Our second heuristic relies, in addition to the graph of commits~$\mathcal{G}_\altcoin$, also on the mapping~$\eventmap$  between 
commits and the events they belong to.
Similarly to {\heuristA}, we look for the earliest non-Bitcoin commit~$C^*_j$ that contains the patch~$C_i$ in its history, the only difference being the meaning of the term ``earliest'':   here, we measure elapsed time with respect to event timestamps, rather than commit timestamps.
According to {\heuristB}, the patch propagation time is defined as the time span between the creation of the patch commit~$C_i$ and the oldest event that references a commit~$C_j$ that has~$C_i$ in its history.

Formally, let~$\EventSet_\altcoin$ denote the set of events pertaining to the altcoin~$\altcoin$ and recorded in the GH archive:
\begin{equation}
	\EventSet_\altcoin := \{ e\ |\ \exists C \in \CommitHistory^\altcoin : \eventmap(C) = e\}.
\end{equation}
According to {\heuristB}, a relevant commit~$C^*_j$ meets the following conditions:
\begin{equation}
	\begin{gathered}
		C^*_j \in \nbcc_\altcoin(C_i) \cap \EventSet_\altcoin\ \land\\
		\eventmap(C^*_j).t \leq \eventmap(C_j).t\ \forall C_j \in  \nbcc_\altcoin(C_i) \cap \EventSet_\altcoin.
	\end{gathered}
\end{equation}
We define the estimated patch propagation time~$\delayB$ as follows:
\begin{equation}
	{\delayB} \gets \eventmap(C^*_j).t - C_i.{t_a}.
\end{equation}

\paragraph{Patch-tag finder ({\heuristC})}
The third heuristic we propose is based on timestamps recorded for relevant tags.
Intuitively, we estimate the patch propagation time as the time between the creation of the Bitcoin patch~$C_i$ and the creation of the first non-Bitcoin tag that links to a commit in~$\altcoin$ that has the patch~$C_i$ in its history.

Formally, we define the concept of ``non-Bitcoin child tag'' analogously to that of non-Bitcoin child commit:
\begin{equation}
	\begin{gathered}
	\tau \in \nbct_\altcoin(C_i) \Longleftrightarrow \tau.h\in \nbcc_\altcoin(C_i)\ \land\\
	\tau \in Tags_\altcoin \land \tau\not\in Tags_{\Bitcoin}.
	\end{gathered}
\end{equation}
According to {\heuristC}, the tag~$\tau^*$ relevant to the patch~$C_i$ is defined as follows:
\begin{equation}
	\begin{gathered}
		\tau^* \in \nbct_\altcoin(C_i)\ \land\ \tau^*.t \leq \tau.t\ \forall \tau \in \nbct_\altcoin(C_i).
	\end{gathered}
\end{equation}
Finally, {\heuristC} estimates the patch propagation delay as follows:
\begin{equation}
	\delayC \gets \tau^*.t - C_i.t_a.
\end{equation}

\paragraph{Comparison between heuristics.}
Assuming successful retrieval of all dangling commits, the patch-event finder ({\heuristA}) can identify every relevant commit~$C^*$ even when multiple rebase operations occurred, by listing all the different versions of~$C^*$ (one per rebase), thus allowing us to select the most accurate commit timestamp.
With other words, {\heuristA} overcomes the problem of retrieving reliable timestamps in the presence of rebasing (c.f.~Section~\ref{sec:git:ops}).
The major limitation of {\heuristA} is that it could under-approximate the patching time, say by~$\Delta$, in case a developer creates the commit locally (or on a dev branch) but waits some time~$\Delta$ before pushing the patch to the repository main branch (e.g., for testing the patched code locally). It further relies on the local clock of the developer which could be skewed or maliciously altered.

Due to its similarity to the graph-based heuristic, {\heuristB} suffers from the same limitation: it could provide a too pessimistic timestamp in case the relevant event has been missed by the event~API. On the other hand, it has the advantage over {\heuristA} that the timestamp cannot be faked or wronged due to a skewed clock.

We expect the tag-based method~{\heuristC} to output the most accurate estimate. Indeed, most users do not compile the latest modifications based on the current version of the main branch (which may be unstable); they are more likely to use released versions of the code, which are marked with tags.

\subsection{Putting all together---{\ours}}

We leverage the aforementioned analysis in devising our tool, {\ours}. To measure the propagation time of patches for a project~$\chi$, {\ours} first has to build the graph $\mathcal{G}_\chi$. Here, \ours{} crawls GH archive for all events pertaining to $\chi$ in order to retrieve all the commits from GitHub's API. Given a patch commit $C_i$
within~$\mathcal{G}_\chi$, {\ours} leverages PCF, PEF and PTF to determine respectively $\Delta_\heuristA$, $\Delta_\heuristB$ and $\Delta_\heuristC$.
The reliance on all three heuristics helps in eliminating possible false positives that may arise due to missing events in the GH archive.
Whenever we obtain different results from the heuristics, 
{\ours} returns the smallest timeframe by default, regardless of which method produced it.
This is the best case for the developers, as some heuristic can output too optimistic values.

To validate the soundness of the proposed heuristics, we use~{\heuristC} as a pessimistic estimate of the propagation time; indeed, a fix may have been introduced before the information provided by the tag, but could not have been produced afterwards. We use this information for cross-validation with the other two approaches, and ensure that the results of the other approaches were consistent with the information constructed from the tag-based analysis. %

Since all three heuristics rely on inspecting Git events, a malfunctioning of the GH archive could harm the accuracy of our tool. More specifically, if the GH archive misses a relevant event (and its corresponding commit), as a consequence our heuristics could over-estimate the patching time.
Another limitation of our heuristics is that they exclusively inspect commits that are either rebased or merged with the same code base: patches introduced with a different code base may not be identified by our tool. For instance, if an altcoin relies on a ``self-made'' patch , and then rebases a few months/years later to the latest version of Bitcoin, our heuristics will point to the time of the rebase.

%% file: methodology.tex
\begin{figure*}
	\centering
	\begin{subfigure}[b]{0.48\textwidth}
		\centering
		\includegraphics[width=\textwidth]{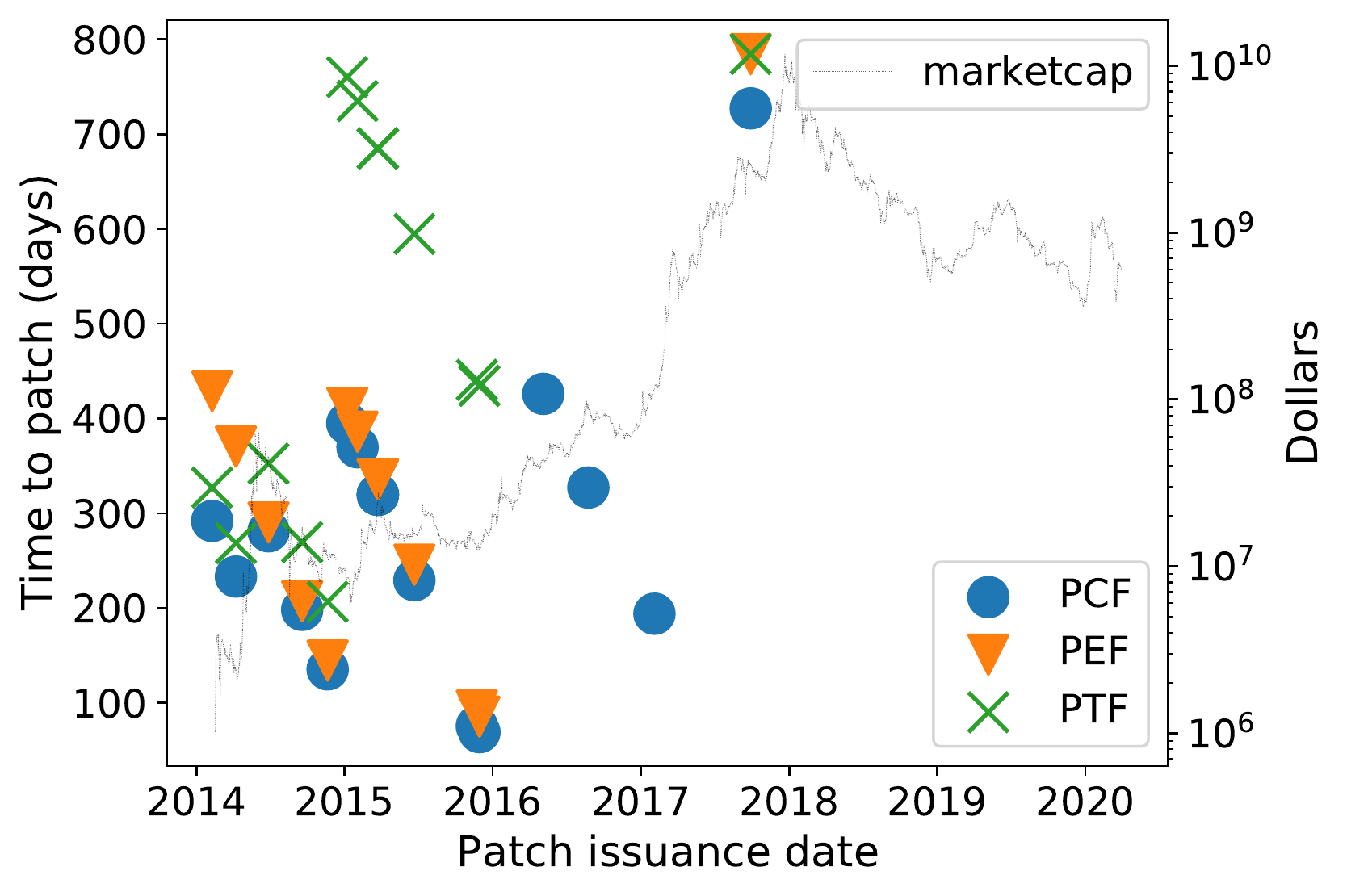}
		\caption{Dash}
		\label{fig:dash}
	\end{subfigure}
	\hfill
	\begin{subfigure}[b]{0.48\textwidth}
		\centering
		\includegraphics[width=\textwidth]{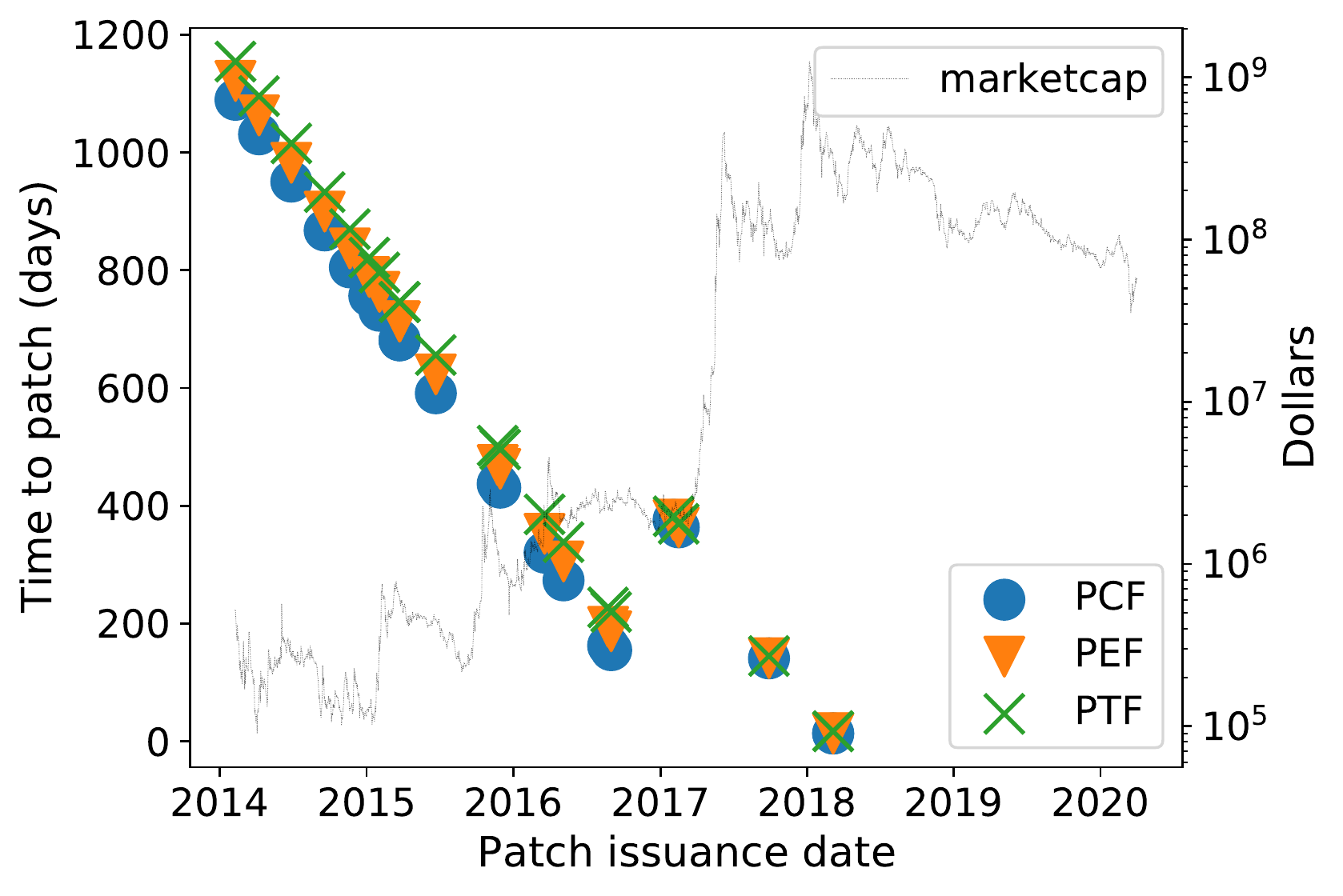}
		\caption{Digibyte}
		\label{fig:digibyte}
	\end{subfigure}
	\hfill
	\begin{subfigure}[b]{0.48\textwidth}
		\centering
		\includegraphics[width=\textwidth]{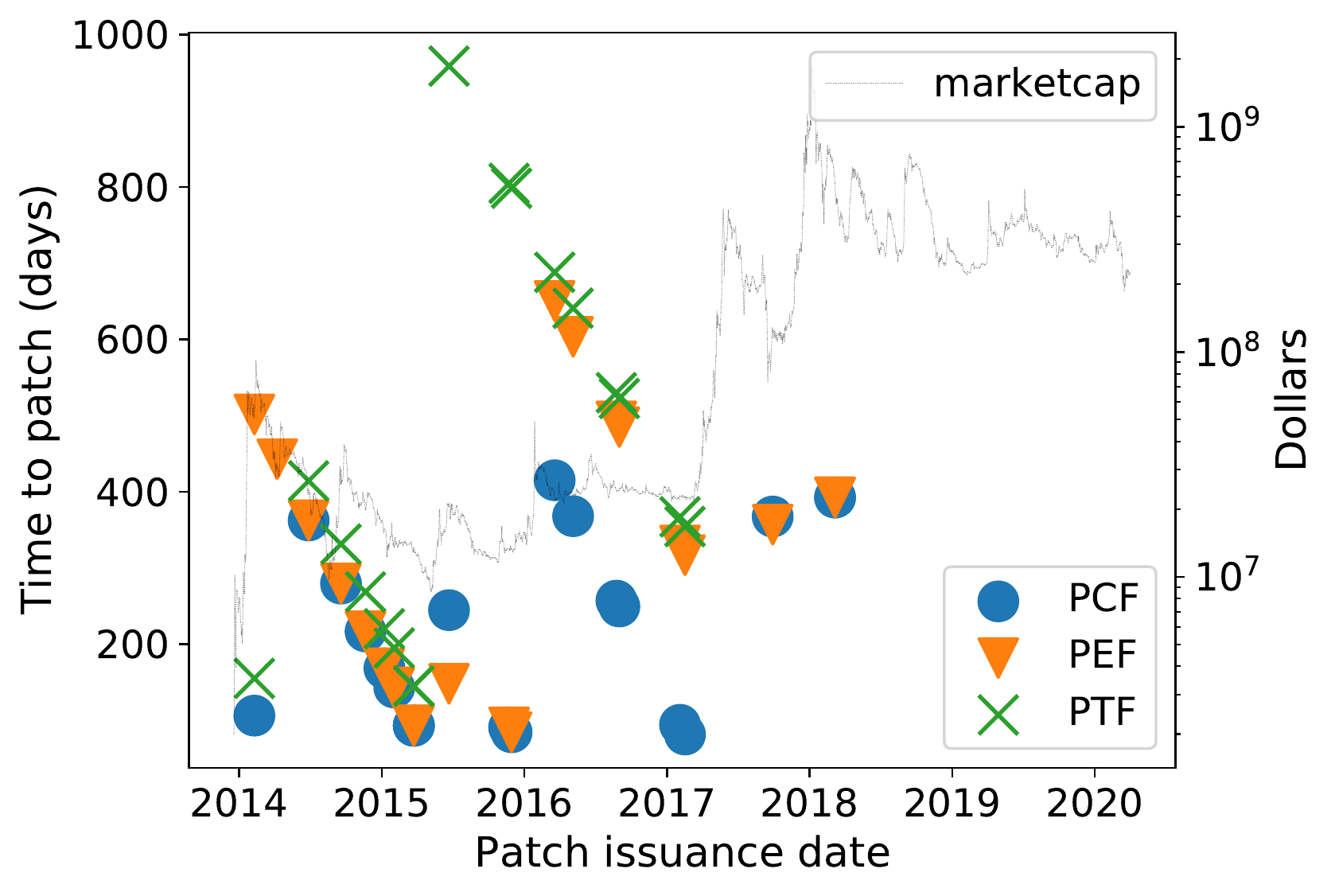}
		\caption{Dogecoin}
		\label{fig:dogecoin}
	\end{subfigure}
	\hfil
	\begin{subfigure}[b]{0.48\textwidth}
		\centering
		\includegraphics[width=\textwidth]{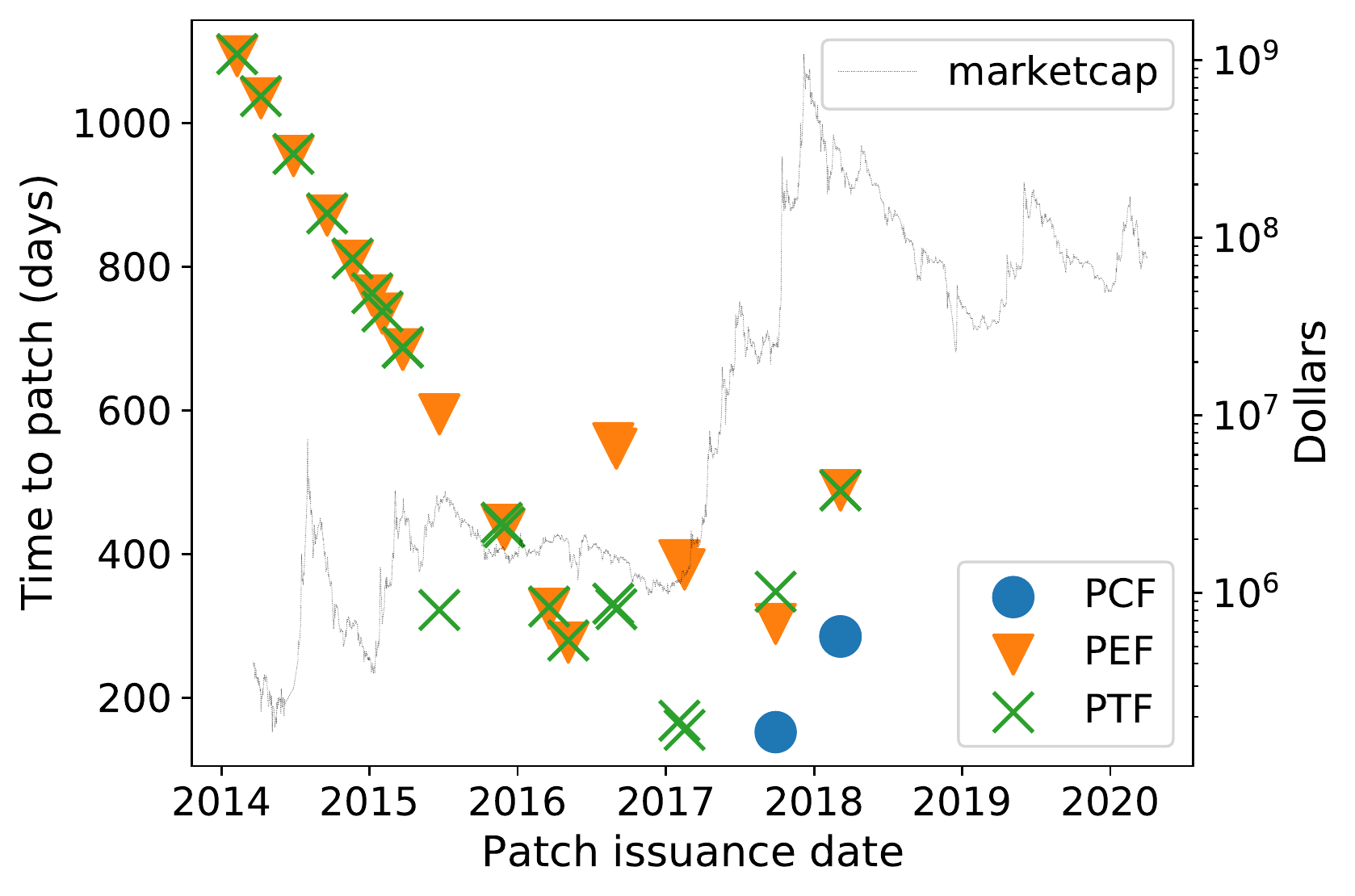}
		\caption{Monacoin}
		\label{fig:monacoin}
	\end{subfigure}
	\caption{Time for a patch issued by Bitcoin-core to be included to the different altcoins. The blue circle, the orange triangle, and the green cross represent respectively the output values given by the graph, the event, and the tag approach. The market capitalisation over time of each coin is plotted as a black dotted line against the values on the right y-axis.}
	\label{fig:timings}
\end{figure*}

\section{Methodology \& Evaluation}
\label{sec:methodology}

In this section, we validate our heuristics by comparing their effectiveness using ground truth data. We then use our heuristics to estimate the time it took to address a large number of patches in Dash, Digibyte, Monacoin, Litecoin, and Dogecoin.


\subsection{Patch selection}
\label{sec:patchselection}

In our evaluation, we restrict our analysis to bugs and patches related to Bitcoin, in particular, how they are propagated through altcoins that are based on the same code base. Since we are interested in patches that are not specific to Bitcoin but relevant to most altcoins, we mainly focus on reported bugs on the peer-to-peer layer as this layer is generally inherited by altcoins (including those that introduce non-negligible modifications to the code base).

We focused on analyzing five altcoins which we selected among existing open-source forks of Bitcoin, namely
Dash~\cite{dash},
initially known for its early adoption in darknet markets, currently worth 2.65 Billion USD;
Digibyte~\cite{digibyte},
a cryptocurrency advertised for its improved functionality and security, currently worth 1.12 Billion USD;
Monacoin~\cite{monacoin},
a cryptocurrency aimed to become a national payment system in Japan, currently worth 0.11 Billion USD;
Litecoin~\cite{litecoin} and
Dogecoin~\cite{dogecoin}, which emerge among the most popular first-generation derivatives of Bitcoin with a market capitalization of~14.82 Billion USD and 40.07 Billion USD respectively.

We then selected a list of 47  Bitcoin commits, 11 representing patches suggested by academic papers~\cite{eclipseattack,tamperingdelivery,doublespending}, 23 representing patches of CVEs, 3 representing Bitcoin improvement proposals, 3 representing CVEs in libraries used by Bitcoin and the remaining 7 representing bugs found on the GitHub repository with tags related to the peer-to-peer network. These patches include the majority of network and peer-to-peer vulnerabilities that were reported in the last decade.

Notice that as the altcoins' software is becoming more mature with each passing year, we have only few datapoints for the last couple of years. This is due to the small number of reported vulnerabilities and CVEs targeting Bitcoin-core recently (most recent CVEs target peripheral software, such as the Lightning Network).

%% file: analysis.tex
\subsection{Validation of our heuristics}
\label{sec:groundtruth}

To validate the effectiveness of our heuristics, we manually searched for publication dates of relevant patches (by investigating corresponding release notes), and we compared these dates with the output of our heuristics for the same vulnerability. Our results, shown in Table~\ref{tab:groundtruth}, confirm that for all the ground-truth data point we found, the actual patching time falls within the interval reported by our heuristics (i.e., between the minimum and maximum estimated propagation time).
As expected, {\heuristC} emerges as the most precise heuristic, especially since release notes are usually part of a new release to which a tag is assigned--- while {\heuristB} is the least reliable as a missed event in GJ archive could cause an over-estimation of the patch propagation time of several months.

\begin{table}[tbp]
    \caption{Ground-truth data to validate the effectiveness of our heuristics.}
    \label{tab:groundtruth}
    \begin{center}
    \scalebox{0.9}{
		\begin{tabular}{lllccc}
		\toprule
		Vulnerability&Altcoin&Time & {\heuristA}&{\heuristB}&{\heuristC}\\
		\midrule
		BIP 65 & Litecoin & 179 days\footnotemark & 159 & 160 & 181\\
		BIP 65 & Dogecoin & 958 days \footnotemark & 244 & 147 & 958 \\
		BIP 66 & Dogecoin & 194 days\footnotemark&142&142&194\\
		CVE-2013-4627 & Litecoin & 33 days\footnotemark & 17&45&18\\
		CVE-2013-4165 & Litecoin & 28 day\footnotemark[5] &10&529&13\\
		\bottomrule
	\end{tabular}
	}
\end{center}
\end{table}
\footnotetext[2]{\url{https://github.com/litecoin-project/litecoin/blob/v0.10.4.0/doc/release-notes-litecoin.md}}
\footnotetext[3]{\url{https://github.com/dogecoin/dogecoin/releases/tag/v1.14-alpha-1}}
\footnotetext[4]{\url{https://github.com/dogecoin/dogecoin/releases?q=BIP66&expanded=true}}
\footnotetext[5]{\url{https://litecoinmirror.wordpress.com/2013/09/04/litecoin-0-8-4-1-release-notes/amp/}
}

\subsection{Analysis}
\label{sec:results}

We now analyse the time it took to patch each of the~47 selected vulnerabilities for the five studied altcoins.

\begin{figure}
	\includegraphics[width=\columnwidth]{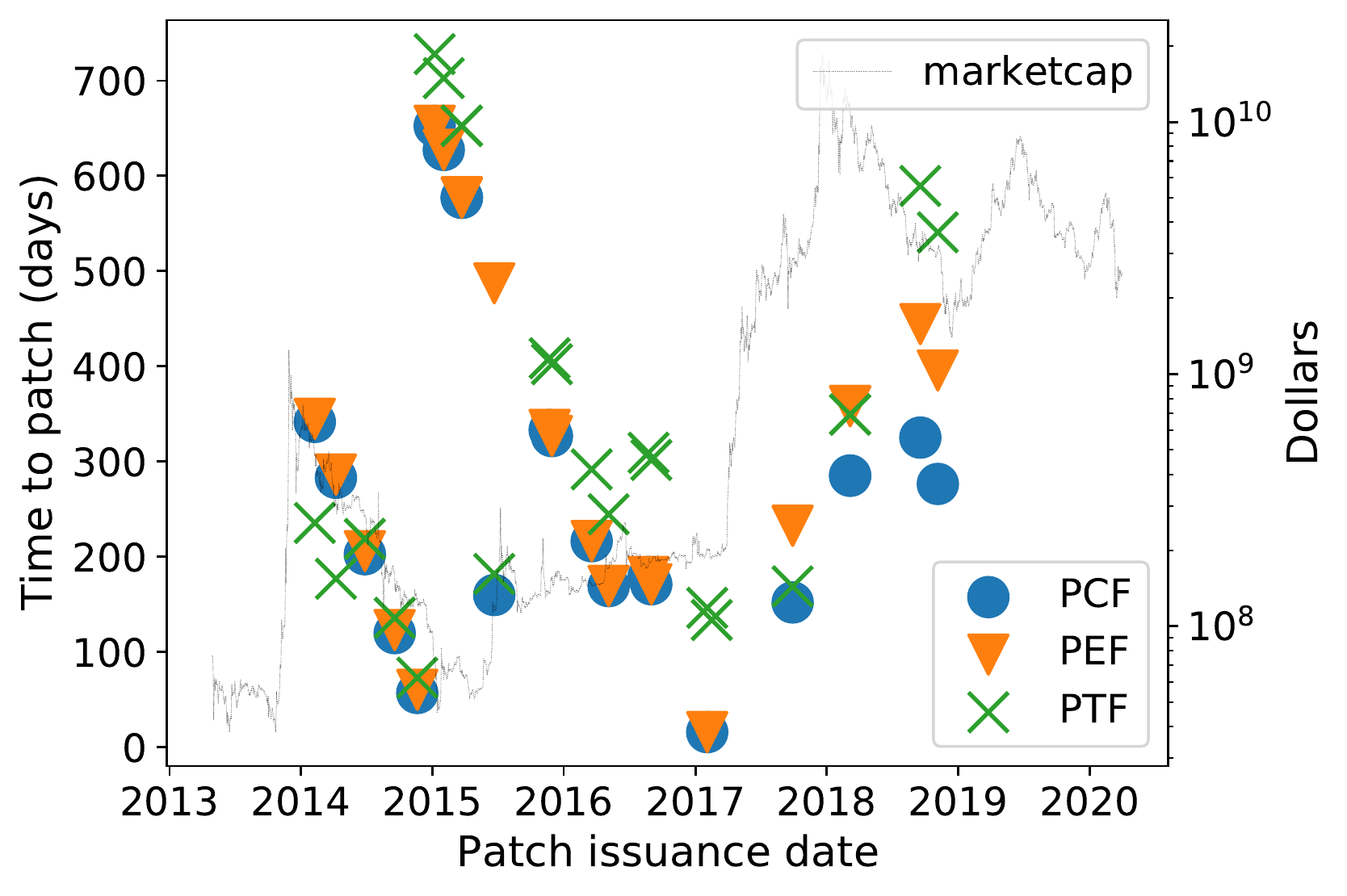}
	\caption{Time for Litecoin to include a patch issued on Bitcoin-core, similar to Figure~\ref{fig:timings}.}
	\label{fig:timings:litecoin}
\end{figure}

As shown in Figures~\ref{fig:timings} and~\ref{fig:timings:litecoin},  our three heuristics provide similar timing estimates, which converge in most cases, thereby supporting the soundness of this approach.
Dash (Figure~\ref{fig:dash}) appears to port patches more quickly, compared to the other blockchains, most of the times with a delay between~200 and 400 days.
Dogecoin and Litecoin instead (Figures~\ref{fig:dogecoin} and~\ref{fig:timings:litecoin}) show more variable patching delays, ranging between~50-600 days, respectively, and 100-500 days on average.
Digibyte and Monacoin (Figures~\ref{fig:digibyte} and~\ref{fig:monacoin}) exhibit an apparent linearly decreasing delay---partly visible also for the other analyzed blockchains. This peculiar behavior suggests that~\texttt{rebase} operations to import the Bitcoin's patches are executed on a regular pace, in a manner that appears to be decoupled from the actual patch release.
This would explain the downward lines in the plots, indicating that groups of patches are actually ported on the corresponding fork at the same time.
To summarize, all five analyzed altcoins apply patches with a delay between several months to a few years.
Based on these results, we conclude that disclosed vulnerabilities remain in the software for several months, as we discuss in detail in Section~\ref{sec:case_studies}.

We include the detailed results of our study in Table~\ref{tab:vulnerability:list}.
Our results show that Bitcoin issues patches to most critical vulnerabilities and CVEs in a prompt manner, often before the publication of vulnerability (i.e., in compliance with the vulnerability disclosure process). That said, there were some cases where Bitcoin took some months to issue a less critical patch reported from the academic community (e.g., forwarding of double spending attempts).

\begin{table}[bp]
    \begin{center}

		\begin{tabular}{c  c  c  c}
			\toprule
			&Patch 1~\cite{patchf59}&Patch 2~\cite{patch5029}&Patch 3~\cite{patch5026}\\
			\midrule
			\textbf{Bitcoin} & \textbf{2014-02-08} & \textbf{2015-11-23} &\textbf{2015-11-29}\\
			\midrule

			Dogecoin & 106 days&90 days &84 days\\

			Monacoin & 1092 &440 days &433 days\\

			Litecoin & 235 days & 333 days &326 days\\

			Digibyte & 1089 days& 437 days&431 days\\

			Dash     & 291 days& 75 days&69 days\\
			\bottomrule
		\end{tabular}
		\caption{Time (from the original Bitcoin patch) to apply the mitigation suggested by Gervais~\emph{et~al.}~\cite{tamperingdelivery}.}
		\label{table:tamperingtiming}
\end{center}
\end{table}

%% file: output.tex
\section{Case Studies}
\label{sec:case_studies}

We now look more closely at two specific vulnerabilities found in Bitcoin:
the first one, disclosed in an academic paper, allows adversaries to tamper with the block delivery of honest users~\cite{tamperingdelivery};
the second one, CVE-2017-18350~\cite{cve18350}, is a reported vulnerability of the Bitcoin software.
We selected these vulnerabilities because they are prominent and relatively recent (disclosure in~2015 and 2017 respectively).
Although both vulnerabilities were patched and released by the Bitcoin-core team, some altcoins---still worth several billions of dollars---applied patches only months or even years after disclosure.

\subsection{Case Study 1: Tampering with the Delivery of Information in Bitcoin~\cite{tamperingdelivery}}

\paragraph{Summary of the vulnerability.}
In order to sustain higher throughput and scalability, Bitcoin implemented a number of optimizations and scalability measures. Gervais~\etal~\cite{tamperingdelivery} show that some of those measures come at odds with the security of the system---effectively allowing adversaries to tamper with the delivery of blocks/transactions to nodes. As a direct outcome of this vulnerability, a resource-constrained adversary would be able to mount a large-scale Denial-of-Service attack on Bitcoin---effectively halting the delivery of all blocks and transactions in the system.

The authors suggested various improvements that resulted in multiple patches, particularly the following three:
\begin{itemize}
    \item Patch 1 - f59d\dots 1392~\cite{patchf59}, penalizing nodes that do not respond to block requests after advertising it.
    \item Patch 2 - 5029\dots 1bff~\cite{patch5029}, enforcing to accept only one INV message per IP address in order to prevent adversaries from filling up the INV log of nodes.
    \item Patch 3 - 5026\dots ca1c~\cite{patch5026}, replacing the invitation message with the full block header, allowing nodes to verify that the received message actually depicts a correctly mined block.
\end{itemize}

\paragraph{Patching time.}
As shown in Table~\ref{table:tamperingtiming}, Dash and Dogecoin took roughly 3 months to port 
these patches from the Bitcoin repository. On the other hand, Monacoin, Litecoin and Digibyte required between 7~months and~3 years.

\subsection{Case Study 2: CVE-2017-18350}

\paragraph{Summary of the vulnerability.}
CVE-2017-18350~\cite{cve18350} is a buffer-overflow vulnerability of the Bitcoin-core software. The vulnerability was located in the proxy support and would enable a malicious proxy server to overwrite the program stack, allowing it to perform remote code execution. However, to be vulnerable, the wallet software needs to be configured to use a malicious proxy, therefore reducing the general risk on the users. Since remote code execution could allow any third party full access to the machine running the node, we deemed that this CVE is of particular interest due to its potential drastic impact.

\begin{table*}[h!]
    \footnotesize
    \caption{\small Summary of the time to patch the vulnerabilities studied in the paper. Here, we use the best (i.e. earliest) result out of the three heuristics. We use a dash (-) when our heuristics could not find the patch in the altcoin, while ``NA'' refers to the case where the vulnerability was discovered and patched prior to the release of the altcoin. All the values are in number of days}
    \label{tab:vulnerability:list}
    \begin{center}
    \begin{tabular}{lllcccccc}
    \hline
    \textbf{Name} & \textbf{Published Date} & \textbf{Description} & \rotatebox{60}{\textbf{Bitcoin}} &     \rotatebox{60}{\textbf{Litecoin}}& \rotatebox{70}{\textbf{Dash}}& \rotatebox{60}{\textbf{Dogecoin}}&     \rotatebox{60}{\textbf{Digibyte}}& \rotatebox{60}{\textbf{Monacoin}}\\
    \hline
	Paper~\cite{eclipseattack} & 2015-08-14 & deterministic random eviction& -143& 19 & 15 & 92 & 681 & 684 \\
	Paper~\cite{eclipseattack} & 2015-08-14 & random selection sha1& -143& 19 & 15 & 92 & 681 & 684 \\
	Paper~\cite{eclipseattack} & 2015-08-14 & random selection sha2& -143& 19 & 15 & 92 & 681 & 684 \\
	Paper~\cite{eclipseattack} & 2015-08-14 & test before evict& 935& 285 & - & 392 & 13 & 285 \\
	Paper~\cite{eclipseattack} & 2015-08-14 & feeler connections& 375& 58 & 327 & 256 & 162 & 165 \\
	Paper~\cite{eclipseattack} & 2015-08-14 & more buckets& -143& 19 & 15 & 92 & 681 & 684 \\
	Paper~\cite{eclipseattack} & 2015-08-14 & more outgoing connections& 1482& - & - & - & - & - \\
	Paper~\cite{tamperingdelivery} & 2015-10-16 & no inv messages& 44& 326 & 69 & 84 & 430 & 433 \\
	Paper~\cite{tamperingdelivery} & 2015-10-16 & filtering inv by ip address& 38& 333 & 75 & 90 & 437 & 440 \\
	Paper~\cite{tamperingdelivery} & 2015-10-16 & penalizing non-responding nodes& -615& 235 & 291 & 106 & 1089 & 1092 \\
	Paper~\cite{doublespending} & 2012-10-18 & forward double spending attempts& 617& 202 & 281 & 362 & 950 & 953 \\
	Vulnerability & - & limit the number of IPs addrman learns from each DNS seeder& 0& 103 & - & 392 & 13 & 103 \\
	Vulnerability & - & ensure tried table collisions eventually get resolved& 0& 281 & - & - & - & - \\
	GitHub bug & - & fixes fee estimate and peers files only when initialized& 0& 119 & 198 & 279 & 867 & 870 \\
	GitHub bug & - & check block header when accepting headers from peers& 0& 56 & 135 & 216 & 804 & 808 \\
	GitHub bug & - & introduce block download timeout& 0& 8 & 87 & 168 & 756 & 759 \\
	GitHub bug & - & fix de-serialization bug where AddrMan is left corrupted& 0& 169 & 426 & 367 & 273 & 276 \\
	GitHub bug & - & dont deserialize nVersion into CNode& 0& 15 & 194 & 94 & 376 & 167 \\
	CVE-2010-5137 & 2010-07-28 & DoS: OP\_LSHIFT crash& 18& NA & NA & NA & NA & NA \\
	CVE-2010-5141 & 2010-07-28 & Theft: OP\_RETURN could be used to spend any output. & 3& NA & NA & NA & NA & NA \\
	CVE-2010-5138 & 2010-07-29 & DoS: Unlimited SigOp DoS & 40& NA & NA & NA & NA & NA \\
	CVE-2010-5139 & 2010-08-15 & Inflation: Combined output overflow& 1& NA & NA & NA & NA & NA \\
	CVE-2010-5140 & 2010-09-29 & DoS: Never confirming transactions & 1& NA & NA & NA & NA & NA \\
	CVE-2011-4447 & 2011-11-11 & Exposure: Wallet non-encryption & 4& 210 & NA & NA & NA & NA \\
	CVE-2012-1909 & 2012-03-07 & Netsplit: Transaction overwriting & -4& 101 & NA & NA & NA & NA \\
	CVE-2012-1910 & 2012-03-17 & Non-thread safe MingW exceptions & -1& 88 & NA & NA & NA & NA \\
	CVE-2012-2459 & 2012-05-14 & Netsplit: Block hash collision (via merkle root)& -14& 43 & NA & NA & NA & NA \\
	CVE-2012-3789 & 2012-06-20 & DoS: (Lack of) orphan txn resource limits& -33& 25 & NA & NA & NA & NA \\
	CVE-2012-4684 & 2012-08-24 & DoS: Network-wide DoS using malleable signatures in alerts& 2& 263 & NA & NA & NA & NA \\
	CVE-2013-2272 & 2013-01-11 & Exposure: Remote discovery of node's wallet addresses& 15& 110 & NA & NA & NA & NA \\
	CVE-2013-2273 & 2013-01-30 & Exposure: Predictable change output& 0& 106 & NA & NA & NA & NA \\
	CVE-2013-3219 & 2013-03-11 & FakeConf: Unenforced block protocol rule& 7& 60 & NA & NA & NA & NA \\
	CVE-2013-3220 & 2013-03-11 & Netsplit: Inconsistent BDB lock limit interactions& 7& 60 & NA & NA & NA & NA \\
	BIP 0034 & 2013-03-25 & FakeConf: block protocol update& -217& 269 & NA & NA & NA & NA \\
	CVE-2013-4627 & 2013-06-01 & DoS: Memory exhaustion with excess tx message data& 62& 17 & NA & NA & NA & NA \\
	CVE-2013-4165 & 2013-07-20 & Theft: Timing leak in RPC authentication& 19& 10 & NA & NA & NA & NA \\
	CVE-2013-5700 & 2013-09-04 & DoS: Remote p2p crash via bloom filters& -15& 0 & NA & NA & NA & NA \\
	CVE-2014-0160 & 2014-04-07 & Remote memory leak via payment protocol& 1& 176 & 233 & 0 & 1030 & 1034 \\
	BIP 66 & 2015-02-13 & FakeConf: Strict DER signatures& -12& 4 & 61 & 142 & 731 & 734 \\
	BIP 65 & 2015-11-12 & FakeConf: OP\_CHECKLOCKTIMEVERIFY& -143& 159 & 229 & 147 & 591 & 322 \\
	CVE-2016-10724 & 2018-07-02 & DoS: Alert memory exhaustion& -836& 216 & - & 414 & 320 & 323 \\
	CVE-2018-17144 & 2018-09-17 & Inflation: Missing check for duplicate inputs& 0& 1 & 1 & - & 155 & 1 \\
	CVE-2017-18350 & 2019-06-22 & Buffer overflow from SOCKS proxy& -632& 151 & 727 & 91 & 140 & 151 \\
	CVE-2018-20586 & 2019-06-22 & Deception: Debug log injection via unauthenticated RPC& -229& 41 & 380 & - & 106 & 244 \\
	CVE-2014-0224 & 2014-06-05 & OpenSSL CVE& 0& 118 & 174 & 0 & 972 & 975 \\
	CVE-2018-12356 & 2018-06-14 & Regex bug& 1& 184 & - & 291 & 50 & 184 \\
	CVE-2019-6250 & 2019-01-13 & Vulnerability in the ZeroMQ libzmq library& 5& 31 & - & - & - & 31 \\
	\hline
	 & & Average&7.53&114.85&188.0&185.17&519.55&503.3\\
	\hline
	& & Number of fixes &47/47&41/42&21/28&23/28&25/28&26/28\\	\hline
    \hline
    \end{tabular}
    \end{center}
\end{table*}

\paragraph{Patching time.}

This vulnerability was discovered on September 21st, 2017 and was patched two days later, on September~23rd. Moreover, the patch was merged with the main branch of the Bitcoin-core repository four days later on September~27th, 2017, and was included in the subsequent releases.

To give enough time to the users for applying the patch, the CVE itself was published only on the June 22nd, 2019. While this patch was applied directly to most of the different altcoins based on the Bitcoin-core software, Dash~\cite{dash} only patched it several months after it was published, on November~19th, 2019. While no attack performed through this vulnerability was reported as far as we are aware, Dash users were seemingly using a vulnerable software with no available update for several months after the disclosure of the vulnerability.

Dogecoin, Digibyte, Monacoin and Litecoin took respectively~91 days, 140 days, 151 days and~151 days to patch this vulnerability after it was discovered. While they all patched it before the vulnerability was disclosed, the software still remained unpatched for several months.

%% file: related-work.tex
\section{Related Work}

The Bitcoin protocol and many of its descendants have been found vulnerable to a wide variety of attacks.
Gervais~\etal~\cite{DBLP:conf/ccs/GervaisKWGRC16} propose a quantitative framework to analyze the security and performance of proof-of-work blockchains with respect to network propagation, block sizes, block-generation intervals, information-propagation mechanism, and the impact of eclipse attacks.
An analysis of Bitcoin's vulnerabilities from a network-security perspective is proposed by Apostolaki~\etal~\cite{DBLP:conf/sp/ApostolakiZV17}, demonstrating partitioning- and delay attacks via hijacking a small number of prefixes.
Security vulnerabilities have been found also for cryptocurrencies other than proof-of-work blockchains.
Kanjalkar~\etal~\cite{DBLP:conf/fc/KanjalkarKLM19} present a resource exhaustion attack affecting various PoS cryptocurrencies whose code base was forked from a variant of Bitcoin, leveraging that the affected cryptocurrencies did not properly validate the proof of stake before allocating resources to peers.
Heilman~\etal~\cite{DBLP:journals/tosc/HeilmanNTLCVD20} demonstrated a collision-finding attack on IOTA's cryptographic hash function Curl-P-27, leading to forging signatures and multi-signatures of valid spending transactions.

Jia~\etal~\cite{DBLP:conf/msr/JiaFXCWYYL20} compare various altcoins based on code similarities and study the correlation between code innovations and market capitalization. Despite focusing on different aspects than security, the authors suggest that code similarity might indicate inherited vulnerabilities. In~\cite{DBLP:conf/blockchain2/HumTTLHLS20}, Hum~\etal~propose a code evolution technique and a clone detection technique to indicate which cryptocurrencies are vulnerable once a vulnerability has been discovered. However, similar to all GitHub parsers, these techniques cannot infer when a given patch has been ported onto an altcoin in case of rebase operations since such timestamps are overwritten by~\texttt{rebase}.

%% file: conclusion.tex
\section{Conclusion}
\label{sec:conclusion}

In this paper, we showed that various altcoins exhibit weaker stability compared to Bitcoin Core. Beyond confirming the folklore result that patch propagation is slow for some altcoin, we introduced a new technique to estimate patch-propagation times (which is non-trivial for GitHub forks), we shed light on the time for altcoins to propagate security patches, and which altcoins are faster in adopting a patch. For instance, CVE-2017-18350 was patched by Dash 5 months after the public release of the CVE.  Moreover, among the five altcoins we analyzed (some of which are worth several billions) Litecoin is the only one to have consistently ported patches within~1 year of their release.

As a by-product, another important purpose of our work is to motivate the need for a proper responsible disclosure of vulnerabilities to all forked chains prior to any publication of the vulnerability. {\ours} offers an effective means for developers to check whether a given patch has been included in relevant forks---before publicly releasing the CVE.